\documentstyle[prl,aps,multicol,epsf]{revtex}
 
\begin{document} 
\draft

\title{Correlation Effects on the Coupled 
Plasmon Modes of a Double Quantum Well} 
\author{N. P. R. Hill \and J. T. Nicholls \and E. H. Linfield \and
M. Pepper \and  D. A. Ritchie \and G. A. C. Jones} 
\address{Cavendish Laboratory,
University of Cambridge, Madingley Road, Cambridge, CB3 OHE, United
Kingdom} 
\author{Ben Yu-Kuang Hu} 
\address{Mikroelektronik Centret, Danmarks Tekniske Universitet,
Bygning 345\o, DK-2800 Lyngby, Denmark}
\author{Karsten  Flensberg}
\address{Dansk Institut for Fundamental Metrologi, Bygning 307,
Anker Engelundsvej 1, DK-2800 Lyngby, Denmark}
\date{\today} 
\maketitle 

\begin{abstract}                                                               
At temperatures comparable to the Fermi temperature,
we have measured a plasmon enhanced Coulomb drag in a GaAs/AlGaAs 
double quantum well electron system. 
This measurement provides a 
probe of the many-body corrections to the coupled plasmon modes,
and we present a detailed comparison between experiment and theory
testing the validity of local field theories. 
Using a perpendicular magnetic field to raise 
the magnetoplasmon energy we  
can induce a crossover to single-particle Coulomb scattering. 
\end{abstract}

\begin{multicols}{2}

Quantum many-body correlations give rise to novel physics,
especially in restricted dimensions where 
interaction effects are greatly enhanced.  
For example, two-dimensional electron gases (2DEGs)
created at a GaAs/AlGaAs heterojunction,
have proven to be fertile systems  
in which correlations reveal themselves in 
a striking multitude of phenomena.   
In these systems, the kinetic energy can be quenched to the extent
that correlations dominate, giving rise to 
effects such as the fractional quantum Hall effect and
Wigner crystallization.
Furthermore, in double-layer electron systems, it has been shown 
that correlations lead to interesting phenomena 
such as new states in the fractional quantum Hall regime.
Collective modes in coupled electron gases
represent another significant area 
for the study of many-body effects in reduced dimensions,
for example, Neilson {\it et al.}\cite{neilson93} 
showed that the low lying acoustic plasmon mode 
of the double-layer system is strongly affected by correlations. 

In zero magnetic field ($B=0$) correlation effects have been investigated 
by both optical\cite{pinczuk89} and 
compressibility\cite{eis94b} measurements.
Transport measurements, on the other hand, while being a good probe
of single-particle behavior, are generally a 
much less sensitive probe of the correlations.
However, one particular type of transport measurement,
Coulomb drag in double-layer electron systems, 
is predicted\cite{flens94} to be 
a sensitive probe of the double-layer plasmon modes.
Since the plasmon modes are affected by correlations, this allows
the unique possibility for a transport measurement to probe many-body 
correlations in a two-dimensional system.

In this letter we present the first transport measurements 
confirming that coupled plasmons can be probed by Coulomb drag.
We compare the experimental data with theoretical results based
on the random phase approximation (RPA) 
with and without local field 
corrections in the Hubbard approximation.\cite{jons76} 
The theory reproduces the data well without any free fitting parameters.
However, the data leaves some open questions
about the applicability of local 
field corrections at elevated temperatures.

In a drag measurement of two closely spaced, 
electrically isolated 2DEGs, a drag voltage $V_{drag}$ is induced 
in one layer when a current $I_{drive}$ 
is passed through the other.\cite{gram91}
Electron-electron interactions 
transfer momentum from the drive layer (with carrier density $n_{drive}$)
to the drag layer (with density $n_{drag}$).
The transresistivity is defined as
\begin{equation}
\rho_t = \frac{V_{drag}}{I_{drive}} \frac{W}{L} , 
\label{e:1}
\end{equation}
where $W$ is the width of the Hall bar, $L$ is the distance
between voltage probes in the drag layer,
and has been calculated\cite{jauho93,zheng93b,flens95b} to be
\begin{eqnarray}
&& \rho_t  = -\frac{\hbar^2}{8\pi^2e^2n_{drag}n_{drive}k_BT} 
\int_{0}^{\infty} dq \int_{0}^{\infty} d\omega\  
q^3\ 
\nonumber\\&&\times
\left|\frac{V(q)}{\epsilon(q,\omega)}\right|^2 
\frac{{\rm Im}\ \chi(q,\omega)_{drive} {\rm Im}\ \chi(q,\omega)_{drag} }
{{\rm sinh}^2(\hbar\omega/2k_BT)},  
\label{e:2}
\end{eqnarray}
where the static interlayer Coulomb interaction is $V(q)$,
$\epsilon(q,\omega)$ is the dielectric function, 
and the charge density fluctuations in a given layer are 
characterized by the polarizability ${\rm Im}\ \chi(q,\omega)$.
At low temperatures and zero magnetic field,
Eq.~(\ref{e:2}) predicts a $\rho_t \sim T^2$ 
temperature dependence,\cite{jauho93}
in approximate agreement with experiment.\cite{gram91} 
Deviations from $T^2$ behavior at low $T$ have 
been attributed to the exchange of phonons.\cite{gram93} 
In a perpendicular magnetic field $B$
the effects of Landau quantization on the drag
have been investigated at low temperatures;\cite{bons96,wu96,hill96} 
at higher temperatures the transresistivity 
varies as $\rho_t \sim B^2 T$.\cite{hill96}

At low temperatures the single-particle excitation (SPE) spectrum
for wavevectors much less than the Fermi wavevector, $q \ll k_F$, 
is bounded by a line of gradient $v_F$ (the Fermi velocity).
The plasmon dispersion curves $\omega_p(q)$
for a double 2DEG system consist of two branches, 
both of which lie higher in energy 
than the SPE spectrum.\cite{sarma81,santoro88,drr96a}
The lower (upper) branch is the acoustic (optic) plasmon,
where the charge density oscillations 
in the two layers are in antiphase (phase).
At $T=0$ the real part of the dielectric function is zero,
${\rm Re}\  \epsilon(q,\omega_p)=0$,
leading to an ``antiscreening'' of the interlayer 
Coulomb interaction for $\omega_p (q)$.
However, at low $T$ there is no coupling between the 
SPEs and the plasmons (${\rm Im}\ \chi(q,\omega_p) = 0$)
and consequently there is no plasmon enhancement of $\rho_t$.
At temperatures of the order of the Fermi temperature, $T \sim T_F$, 
the SPE spectrum is sufficiently broadened to give
a strong contribution to $\rho_t$ from the acoustic plasmon pole,
through both ${\rm Im}\ \chi(q,\omega_p)$ and $\epsilon(q,\omega_p)$.
The optic plasmon has a higher energy, 
and therefore makes a smaller contribution
to the integral in Eq.~(\ref{e:2}). 
Detailed calculations\cite{flens94,flens95a} 
show a plasmon enhancement of the scaled transresistivity
$\rho_t T^{-2}$ around $0.2~T_F$, 
which peaks close to $0.5~T_F$.  
For $T>0.5~T_F$ the strong coupling between the plasmons 
and the SPEs causes Landau damping of the two modes 
and the plasmon enhancement of the drag 
diminishes.\cite{flens94,flens95a} 

Samples A and B were fabricated from two similar 
wafers grown by molecular beam epitaxy, 
and consisted of two 200~\AA \ wide modulation-doped
GaAs quantum wells separated by a 300~\AA \ Al$_{0.67}$Ga$_{0.33}$As barrier.  
The resulting center-to-center separation of the two 2DEGs is $d=500$~\AA.
Patterned back-gates were defined in a buried $n^+$ GaAs layer 
using {\it in situ} focused ion beam lithography.\cite{lin93} 
A Hall bar mesa ($W=67~\mu$m and $L=500~\mu$m),
NiCr:Au front-gates, and AuGeNi Ohmic contacts to the back-gates and 2DEGs
were fabricated by optical lithography.  
Independent contacts to the two
layers were then formed using 
a selective depletion technique.\cite{eis90,brown94b}
Surface gates and back-gates, extending over the active
area of the Hall bar, were used to control independently the carrier
densities of the individual layers, 
which were determined by four-terminal
Shubnikov-de Haas and low field Hall measurements. 

The zero field measurements in Figs.~\ref{f:1} and \ref{f:2} 
were obtained from sample A
which has an interlayer resistance $\sim100$~M$\Omega$. 
Upon application of a magnetic field 
the resistive voltage drop in the drive layer increases,
and there is a corresponding 
increase in the leakage current between the layers.
Sample B has an interlayer resistance greater than $1$~G$\Omega$,
and provided the magnetotransport measurements shown in Fig.~\ref{f:3}.
The zero field measurements presented for sample A 
were confirmed with sample B.
Sample A has as-grown carrier densities of
$3.3 \times 10^{11}$~cm$^{-2}$ 
and $2.3 \times 10^{11}$~cm$^{-2}$ with low
temperature mobilities of $9.0 \times 10^{5}$~cm$^{2}$/Vs and
$1.3 \times 10^{5}$~cm$^{2}$/Vs in the upper and lower 2DEG respectively.  
The corresponding values for sample B are 
$3.1 \times 10^{11}$~cm$^{-2}$, $2.2 \times 10^{11}$~cm$^{-2}$ 
and $6.5 \times 10^{5}$~cm$^{2}$/Vs, 
$7.6 \times 10^{5}$~cm$^{2}$/Vs.
Samples A and B show quantitatively similar drag results.

The drag measurements were made using a circuit and ac lockin techniques,
which have been described elsewhere.\cite{gram91}
The measured transresistivity was identical when 
the roles of the upper and lower 2DEG were reversed,
and $\rho_t$ scaled with $L$ according to Eq.(~\ref{e:1}). 
The drag measurements remained unchanged when
the position of the earth on the drag 
layer was switched between the voltage probes,
demonstrating that there are no interlayer leakage effects.\cite{gram91}

Figure~\ref{f:1} shows the scaled transresistivity
$\rho_t T^{-2}$ versus $T/T_F$ for matched carrier
densities $n = 1.37$ to $2.66 \times 10^{11}$cm$^{-2}$.
For all densities the traces show an upturn close to $0.2~T_F$
with a maximum close to $0.5~T_F$,
in good qualitative agreement with theoretical predictions.
The dashed lines in Fig.~\ref{f:1} show
the scaled transresistivity based on RPA calculations\cite{flens94}
of the coupled plasmon dispersion relations using the 
carrier densities and structural parameters of our samples.
These calculations evaluate the Coulomb coupling between the 
layers and do not include the phonon 
exchange that is measured below 0.1~$T_F$.

The calculations, while reproducing the overall shape 
of the drag, show discrepancies with the experiment. 
The temperature required to excite the plasmon is lower 
in the experimental traces and,
over the most of the temperature range,
the magnitude of the drag is larger than the prediction. 
The maximum in $\rho_t T^{-2}$ occurs at a lower temperature 
than the calculation and the decline of the enhancement 
at higher temperatures is more pronounced,
indicating that the RPA underestimates the Landau 
damping effect in this regime.
Both features suggest that the approximations of the RPA
lead to a plasmon dispersion relation above its true value, 
and that a more sophisticated consideration of 
the collective excitations is required.

The solid lines in Fig.~\ref{f:1}
show calculations of the scaled transresistivity,
where intralayer exchange interactions are included 
in the Hubbard approximation. 
The local field calculations 
show qualitatively the same results as the RPA;
however, the inclusion of many-body correlations 
lowers the plasmon energy,
thereby increasing the plasmon contribution to the 
drag and lowering the temperature 
required to excite the acoustic plasmon. 
The Hubbard approximation provides a better fit to the 
data and the temperature required to excite the plasmon
is better described by the theoretical curves. 
The position of the predicted maximum in $\rho_t T^{-2}$ 
moves to lower temperature due to the increased influence 
of Landau damping and the fit to 
the experimental data points improves. 
However, there is still a significant 
discrepancy in the magnitude of the drag. 
Recent theoretical work\cite{swier95} has 
emphasized the role of many-body correlations in a 
Singwi-Tosi-Land-Sj\"{o}lander (STLS) description 
of the drag and gives similar results to the Hubbard 
approximation.\cite{STLS96}
Both calculations use zero temperature local field corrections to 
describe the correlations; in contrast,
the measurements are carried out at a significant fraction of the 
Fermi temperature, possibly leading to considerable 
modifications of the local fields. 
In fact, the zero temperature local field corrections
overestimate the correlation effects at high temperatures.
Hence improved finite temperature
calculations are needed in order to learn 
more about the correlations in coupled electron
gases at elevated temperatures.

Figure~\ref{f:2} shows the scaled transresistivity 
as a function of the relative carrier density, 
$n_{drag}/n_{drive}$,
when the drive layer was fixed at $2.66 \times 10^{11}$cm$^{-2}$. 
The plasmon enhancement described by Eq.(~\ref{e:2})
depends on the product 
${\rm Im}\ \chi(q,\omega_p)_{drive} {\rm Im}\ \chi(q,\omega_p)_{drag}$,
which has a maximum when the boundaries 
of the SPE in each layer are equidistant from the 
acoustic plasmon dispersion curve; therefore a peak in $\rho_t$
is expected\cite{flens94} at matched Fermi velocities,
$n_{drag} = n_{drive}$. 
At $T = 0.16~T_F$ the drag 
shows a monotonic decrease with $n_{drag}/n_{drive}$,
characteristic of single-particle Coulomb scattering; 
the maximum at $n_{drag} = n_{drive}$
develops with increasing temperature as the SPE spectrum is 
smeared into the plasmon dispersion curve. 
The theoretical curves in the RPA and 
Hubbard approximation also show the evolution of the plasmon
signature at matched $n$, but with the same 
discrepancy in magnitude that is evident in Fig.~\ref{f:1}. 
When the temperature is lowered below $T=0.16~T_F$ 
the phonon interaction becomes the 
dominant interlayer scattering mechanism and the calculated 
contribution from Coulomb coupling does not describe the data. 
Instead the experimental traces show the 
development of the $\rho_t$ maximum at matched $n$ 
that is characteristic of phonon exchange.\cite{gram93}
In summary, Fig.~\ref{f:2} demonstrates the crossover between 
the three different interlayer scattering mechanisms:
phonon exchange for $T < 0.1~T_F$,
single-particle Coulomb scattering for $T \approx 0.16~T_F$,
and plasmon enhancement for $T > 0.2~T_F$.
A further transition back to single-particle Coulomb scattering 
is expected at even higher temperatures 
when the plasmon modes are heavily Landau damped. 

In Fig.~\ref{f:3} we investigate the effect of a magnetic field 
on the plasmon enhancement at high temperatures. 
The magnetic field is always such that the Landau 
level (LL) structure is thermally smeared 
and LL structure is not seen in the magnetoresistance 
traces of the individual layers. 
The scaled transresistivity
is measured as a function of $n_{drag}/n_{drive}$ 
at $T=40$~K for increasing $B$,
when the drive layer density 
was fixed at $2.23 \times 10^{11}$~cm~$^{-2}$.
We also observe an overall increase of $\rho_t$ in a magnetic field, 
which has been previously been reported.\cite{hill96}. 
At low fields the $\rho_t$ maximum at $n_{drag}/n_{drive}=1$ 
has the same origin as the plasmon enhanced 
peak identified in Fig.~\ref{f:2}.

In a magnetic field the magnetoplasmon dispersion relation,
$\omega(B,q) = \sqrt{\omega_c^2+\omega^2_{p}(q)}$,
is pushed to higher energies than 
the dispersion $\omega_{p}(q)$ at $B=0$.
Therefore when $\hbar\omega_c \gtrsim k_BT$
the magnetoplasmon cannot be so easily excited
and consequently the plasmon contribution decreases.
Figure~\ref{f:3} shows the matched carrier density peak,
a signature of plasmon enhancement at high temperatures,
becoming weaker as $B$ is increased. 
The matched $n$ peak disappears at $B \approx 2$~T,
the magnetic field at which $\hbar \omega_c$ corresponds
to the $k_B T$ at 40~K.
At $B=2.2$~T the transresistivity monotonically decreases
with increasing $n_{drag}/n_{drive}$,
displaying behavior similar to that observed at 
$T \approx 0.16~T_F$ in Fig.~\ref{f:2}, 
which has been identified as single-particle Coulomb scattering.
The magnetic field in Fig.~\ref{f:3} has an opposite effect 
to that of temperature in Fig.~\ref{f:2}. 
By increasing the temperature at zero field we can induce a crossover 
from single-particle Coulomb scattering to plasmon enhancement,
as the SPEs are smeared into the plasmon dispersion curves. 
The reverse transition can be induced with a magnetic field, 
lifting the magnetoplasmon energy further above the SPE boundary. 

In conclusion, we have measured a plasmon enhancement of 
the Coulomb drag in a double quantum well system. 
We have observed a crossover in both temperature and magnetic field 
from single-particle behavior to coupled plasmon enhancement.
A detailed comparison between experiment and theory
shows the importance of many-body corrections to the coupled plasmon modes.
The results presented in this letter should provide a catalyst 
for further theoretical work into the applicability of local 
field corrections at elevated temperatures.
The plasmon enhancement of Coulomb drag is 
an ideal testing ground for such research.

We wish to thank the Engineering and Physical Sciences 
Research Council (UK) for supporting this work.  
JTN acknowledges an EPSRC Advanced Fellowship, 
and DAR acknowledges support from the Toshiba Cambridge Research Centre.  
We thank Antti-Pekka Jauho for useful discussions.

\end{multicols}

\begin{figure} 
\caption{The scaled transresistivity $\rho_t T^{-2}$
versus the reduced temperature $T/T_F$ of sample A,
for matched carrier densities
$n=1.37, 1.80, 2.23$, and $2.66 \times 10^{11}$~cm$^{-2}$.
The dashed (solid) lines are RPA (Hubbard) calculations of 
the scaled transresistivity. 
The Hubbard approximation with zero temperature local field corrections
improves the agreement between theory and experiment
in the low temperature region, but overestimates the 
correlations at higher temperatures.} 
\label{f:1} 
\end{figure}

\begin{figure} 
\caption{The scaled transresistivity $\rho_t n_{drag}/n_{drive}$ 
of sample A as a function of relative carrier density $n_{drag}/n_{drive}$,
for different reduced temperatures. 
The density of the drive layer was fixed at 
$n_{drive}= 2.66 \times 10^{11}$~cm$^{-2}$ ($T_F =110$~K).
The dashed (solid) lines are RPA (Hubbard) 
calculations of the scaled transresistivity.
Note that the low temperature traces are dominated by the
phonon exchange mechanism and are not expected to be reproduced
by theory. As in Fig.~\ref{f:1} we see that the Hubbard approximation
overestimates the correlation effects at high temperatures,
whereas the intermediate temperature regime is well reproduced by theory.} 
\label{f:2}
\end{figure}

\begin{figure}
\caption{The scaled transresistivity $\rho_t n_{drag}/n_{drive}$ 
of sample B as a function of the relative carrier density at $T=40$~K,
for $B=0$, 0.6, 1.2, 1.8, and 2.2~T.
The density of the drive layer was fixed at 
$n_{drive}= 2.23 \times 10^{11}$~cm$^{-2}$ ($T_F =93$~K).
The figure shows how the plasmon peak disappears as the 
magnetoplasmons are pushed to higher energies and the
coupled plasmon can not be thermally excited.}
\label{f:3} 
\end{figure}

\end{document}